\documentclass[twocolumn,aps,prb]{revtex4-1}
\pdfoutput=1
\usepackage{amssymb}
\usepackage[latin1]{inputenc}
\usepackage{amsmath}
\usepackage{graphicx}
\usepackage{color}
\usepackage[english]{babel}

\makeatletter
\begin{document}
%\setpagewiselinenumbers
%\modulolinenumbers[1]
%\linenumbers

\title{Power and linewidth of propagating and localized modes\\ in nanocontact spin-torque oscillators}

\author{Stefano Bonetti}
\email{bonetti@slac.stanford.edu}
\affiliation{Materials Physics, KTH - Royal Institute of Technology, Stockholm, Sweden}
\affiliation{Department of Physics, Stanford University, Stanford, CA, USA}
\author{Vito Puliafito}
\affiliation{Department of Matter Physics and Electronic Engineering, University of Messina, Messina, Italy}
\author{Giancarlo Consolo}
\affiliation{Department of Sciences for Engineering and Architecture, University of Messina, Messina,Italy}
\author{Vasyl S. Tiberkevich}
\author{Andrei N. Slavin}
\affiliation{Department of Physics, Oakland University, Rochester, MI, USA}
\author{Johan \r{A}kerman}
\affiliation{Department of Physics, University of Gothenburg, Gothenburg, Sweden}
\affiliation{Materials Physics, KTH - Royal Institute of Technology, Stockholm, Sweden}

\begin{abstract}
%Measurements of the integrated power ($P$) and linewidth ($\Delta f$) of microwave signals originating from localized and propagating spin-wave modes in nanocontact spin-torque oscillators (NC-STOs) are presented as functions of the applied field angle $\theta_e$ (w.r.t. to the film plane). For the propagating mode, we find that $P$ increases with $\theta_e$ in an essentially monotonic fashion. For the localized mode, we observe a broad maximum in $P$ at $\theta_e\approx25$--$50$ deg, then an exponential decrease down to the critical angle $\theta_c\approx58$ deg, above which this mode no longer exists. These observations are qualitatively reproduced by micromagnetic simulations, and are related to the precession orbit of the magnetization. The experimentally determined $\Delta f$ agrees well with the analytical theory when only the propagating mode is excited. However, for $\theta_e<\theta_c$, i.e. when both modes are excited, the measured $\Delta f$ is more than an order of magnitude larger than the theoretical predictions. We are able to explain the linewidth broadening in this case assuming partially coherent mode hopping between the two spin-wave modes, in which the phase of the oscillation is randomized only at the transition between the two modes. An average phase slip of $\Delta\phi\approx75$ deg is found to describe the experimental data well.

Integrated power and linewidth of a propagating and a self-localized spin wave modes excited by spin-polarized current in an obliquely magnetized magnetic nanocontact are studied experimentally as functions of the angle $\theta_e$ between the external bias magnetic field and the nanocontact plane. It is found that the power of the propagating mode monotonically increases with $\theta_e$, while the power of the self-localized mode has a broad maximum near $\theta_e = 40$ deg, and exponentially vanishes near the critical angle $\theta_e = 58$ deg, at which the localized mode disappears. The linewidth of the propagating mode in the interval of angles $58<\theta_e<90$ deg , where only this mode is excited, is adequtely described by the existing theory, while in the angular interval where both modes can exist the observed linewidth of both modes is substantially broadened due to the telegraph switching between the modes. Numetical simulations and an approximate analytical model give good semi-quantitative description of the observed results.
\end{abstract}

\maketitle

\section{Introduction}
Current-induced spin-wave dynamics \cite{Slonczewski1996,Berger1996} has attracted great attention in recent years, both from applied and fundamental points of view.\cite{Kiselev2003,Kiselev2004PRL,Rippard2004a,bertotti2005prl,Krivorotov2005aScience,Rippard2006PRB,Houssameddine2007,Pribiag2007,berkov:144414,Krivorotov2007,DeacNatPhys08,Thadani08,Houssameddine2008,bonetti:102507,lehndorff:054412} In terms of applications, the possibility of realizing nanoscale microwave oscillators---so-called spin-torque oscillators (STOs)---is very appealing. From a fundamental perspective, this topic offers a new field of investigation in which spin-wave dynamics can be studied at reduced dimensions.

In an STO, a high-density spin-polarized direct current ($\sim~10^8$ A/cm$^2$) transfers part of its spin angular momentum to a ferromagnetic (FM) thin film via a spin-transfer torque effect.\cite{Slonczewski1996,Berger1996} Under certain conditions of the applied magnetic field and current, a sustained precession of the magnetization in the ferromagnetic film can be achieved. Typically, a pseudo spin-valve stack is needed to both spin-polarize the current through one of the magnetic layers (referred to as the ``fixed'' layer), and to be able to read the time-varying giant magnetoresistance (GMR) signal arising from the magnetization precession of the other layer (thought of as a ``free'' layer). Indeed, the way signals are detected in these devices is as an oscillating voltage, which arises because of Ohm's law, which in this case can be written as $V(t)=R(t)I_{dc}$.

STOs can be realized in two geometries: nanocontacts (NCs), where the only nanosized feature is the nonmagnetic electrode through which the high density current is injected, and nanopillars, in which the magnetic layers, in whole or in part, are nanosized. In this work, we will consider the case of NC-STOs. These devices are particularly interesting from a fundamental perspective, because they represent an ideal system in which the characteristics of current-induced spin waves can be studied. In fact, in such devices the spin-valve stack is patterned into a larger mesa with a lateral size of several microns. This allows for the spin waves generated at the NC to be damped out before reaching the edges of the mesa, and thus avoiding eventual reflections, which may induce more complex effects in the dynamics.

A few years after the prediction of the feasibility of current-induced spin-wave excitations, Slonczewski proposed that the spin waves generated in a \emph{perpendicularly} magnetized NC are propagating spin waves with wavevector inversely proportional to the NC radius.\cite{Slonczewski1999} While his calculations described this kind of excitation well, they were inadequate for describing the case of \emph{in-plane} magnetized NCs. A theoretical solution was proposed several years later by two of the authors of this article,\cite{Slavin2005PRL} who demonstrated that experimental results could be better explained in terms of a self-localized spin-wave soliton---the so-called spin-wave bullet.

The first studies of spin-wave dynamics in NC-STOs under the effect of oblique magnetic fields were first presented in Refs.~\onlinecite{Rippard2004a,Rippard2006PRB}. However, it was with our previous work \cite{PhysRevLett.105.217204} that it became possible to obtain definitive experimental confirmation of the two theoretical approaches. We found that the localized Slavin-Tiberkevich mode only exists below a certain critical angle $\theta_c$, while the propagating Slonczewski mode can be excited at all angles $\theta_e$. The observed excitation frequencies, threshold characteristics, and frequency agility with current agree with theoretical calculations and micromagnetic simulations.\cite{gerhart:024437,consolo:014420} Furthermore, the very recent observations of spin-torque induced spin waves by means of micro-focused Brillouin light scattering strengthen these conclusions in a direct way, proving that it is the direction of the external magnetic field that determines the character (localized or propagating) of the excited spin waves.\cite{Demidov2010,Madami2011}

Here we investigate both experimentally and by means of micromagnetic simulations the dependence of the power of the two modes on the applied magnetic field angle. The qualitative features of these two cases are remarkably different, and given the agreement between experiments and simulations, we are able to infer details of the magnetization dynamics. Finally, using the analytical model that describes thermal effects in nonlinear auto-oscillators, we are able to describe the experimental angular behavior of the linewidth when only one mode is excited. When two modes are excited, the analytical model underestimates the value of the linewidth by more than one order of magnitude, and we introduce a new model in order to describe this situation.

\section{Experimental and simulation details}
\begin{figure}[t!]
\centering
\includegraphics[trim=5cm 9cm 5cm 9cm clip=true, scale=0.62]{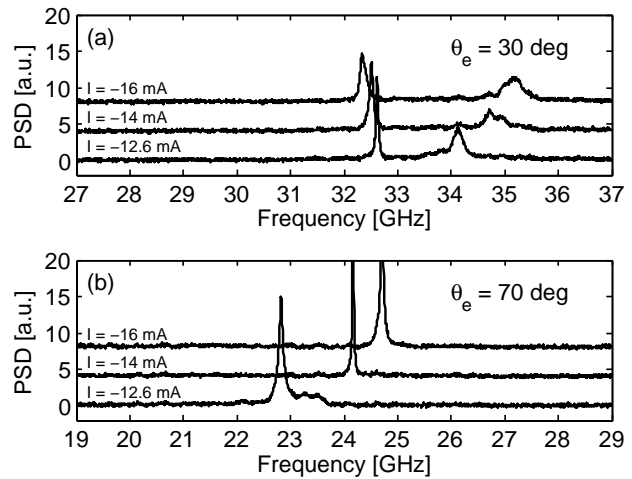}
\caption{Power spectral density at $I_{dc}=$ -12.6, -14, and -16 mA, for two different directions of the external magnetic field. (a) For $\theta_e=30$ deg, two distinct peaks can be identified: a red-shifted lower frequency peak, identified as a localized soliton mode, and a blue-shifted higher frequency peak, identified as the propagating mode. (b) At $\theta_e=70$ deg, only the propagating mode is observed.}
\label{fig:PSDvs_freq}
\end{figure}

In this work, the STO under investigation comprises a $d=40$ nm circular NC on top of a pseudo spin-valve mesa composed of a 20 nm thick Co$_{80}$Fe$_{20}$  ``fixed'' magnetic layer, a 6 nm thick Cu spacer layer, and a 4.5 nm thick Ni$_{80}$Fe$_{20}$ ``free'' magnetic layer. Further details of the material stack can be found in Ref.~\onlinecite{Mancoff2006}. On top of the STO, a coplanar waveguide (CPW) is defined to allow for both injection, through a bias-T, of a current $|I_{dc}|=5$--$18$ mA flowing from the biased free layer to the grounded fixed layer, and for extraction of the microwave signal generated by the device. The external contact with  the CPW consists of nonmagnetic ground-signal-ground probes (dc--40 GHz). A constant magnetic field $\mu_0H_e=1.1$ T %, where $\mu_0$ is the permittivity of free space, 
is applied to the sample. During operation, a current $|I_{dc}|=5$--$18$ mA flows from the biased free layer to the grounded fixed layer, injected via a bias-T.  
The microwave signal is amplified +22 dB and guided into a spectrum analyzer. Details of the measurement setup are given elsewhere.\cite{bonetti:102507} We calibrated our transmission line, however, due to the wide frequency range investigated ($\sim20$ GHz). We are aware that the computed power will not be as accurate as in the case of our recent work,\cite{PhysRevB.81.140408} where the frequency range was much smaller ($\sim500$ MHz). We therefore expect a 3 dB error in power (i.e. a factor 2) to be unavoidable in the present measurements. Nevertheless, given the variation of the STO power over several order of magnitudes as the magnetic field angle is varied, the measurements are meaningful and sufficiently accurate in this context.

In order to compare our experimental results with micromagnetic simulations, we used the data from Ref.~\onlinecite{consolo:014420}. The sample geometry in that case was very similar, although the applied field was slightly different. Since we are only interested in a qualitative understanding of the results obtained by using this experimental setup, a new full-scale ad hoc numerical study was not performed. Details of the micromagnetic framework can be found in Ref.~\onlinecite{consolo:014420}.

\begin{figure}[t!]
\centering
\includegraphics[trim=5cm 7.5cm 5cm 7.5cm clip=true, scale=0.62]{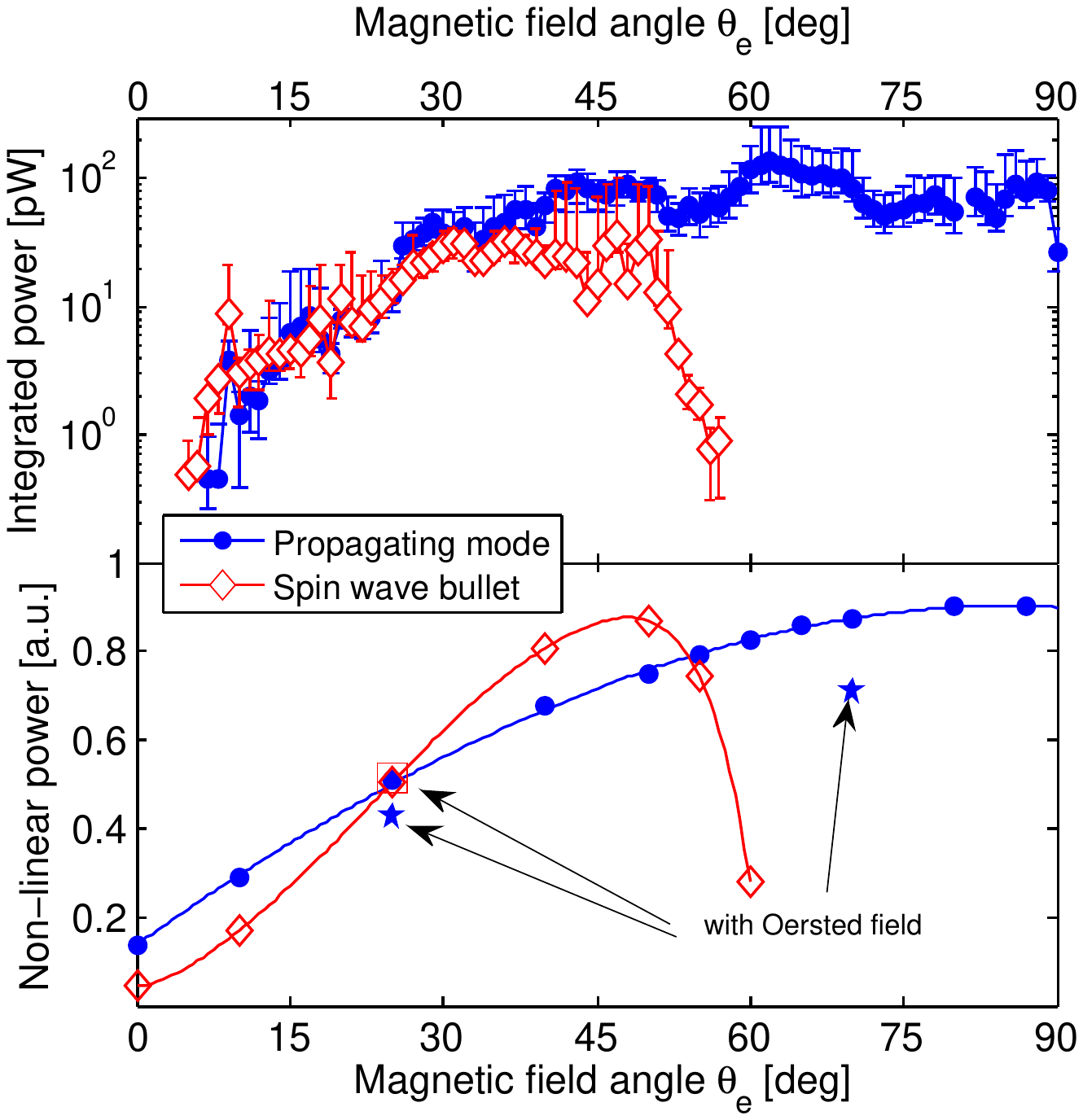}
\caption{(Color online) (a) Averaged experimentally measured integrated power of both modes as a function of angle of applied magnetic field. The asymmetric error bars at each data point represent the upper and lower range of the data set, which is made up of the integrated power at different current values. (b) Non-linear power of both modes computed from micromagnetic simulations using Eq.~(\ref{eq:P_NL}).}
\label{fig:intpow_vs_theta}
\end{figure}

\section{Results}
Fig. \ref{fig:PSDvs_freq} summarizes the results of our previous work.\cite{PhysRevLett.105.217204} In particular, it shows that for a certain angle $\theta_e<\theta_c$ ($\theta_c\approx55$ deg), two spin-wave modes with opposite frequency agility can coexist. Only one mode exists for $\theta_e>\theta_c$. Comparison with theory and micromagnetic simulations allowed us to identify the mode at lower frequency as a localized mode of solitonic character (a spin-wave ``bullet''), and the high frequency mode as a propagating mode.\cite{PhysRevLett.105.217204}

Fig. \ref{fig:intpow_vs_theta}(a) shows the measured integrated power of the two modes as a function of applied magnetic field angle. The integrated power is calculated by averaging the power at each current above the threshold current.\cite{Krivorotov2007} The threshold current is obtained by the procedure illustrated in our previous work.\cite{PhysRevLett.105.217204} Error bars indicate the range over which the power varies at different bias currents. The propagating mode shows an approximately monotonic increase in emitted power as out-of-plane angles are approached. The localized mode, on the other hand, shows a monotonic increase in power up to an angle of $\theta_e\approx30$ deg, followed by a plateau up to $\theta_e\approx45$ deg, and then a rapid decrease down to the critical angle, above which the mode is no longer excited.

For micromagnetic simulations, the power was computed employing the definition of nonlinear oscillation power developed in Ref.~\onlinecite{PhysRevB.81.184411}, and by then performing an analogous averaging over the current values. In Fig. \ref{fig:intpow_vs_theta}(b), we show the power computed by means of this technique, which will be discussed in more detail in the next section. It should be noticed that the power of the propagating mode increases monotonically with the applied field angle, while the power of the localized mode increases up to the maximum at $\theta_e\approx50$ deg, and then undergoes a decrease down to the critical angle.

\begin{figure}[t!]
\centering
\includegraphics[trim=5cm 9cm 5cm 8.5cm clip=true, scale=0.65]{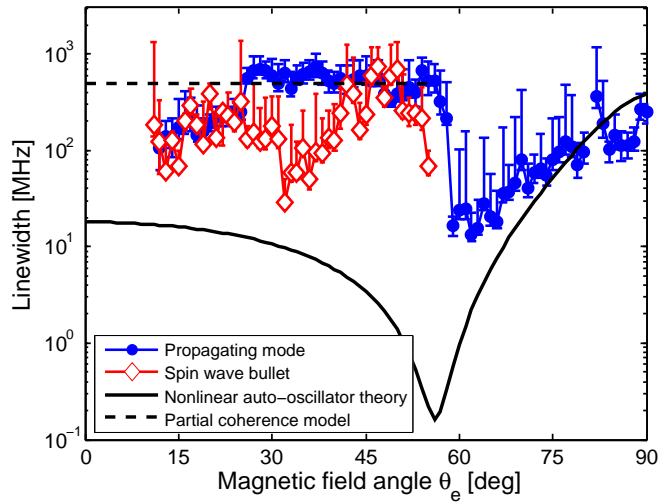}
\caption{(Color online) Linewidth of both modes as a function of angle of applied magnetic field. The continuous line is the linewidth calculated using the analytical model of Ref.~\onlinecite{Kim2008PRL100a}.}
\label{fig:linewidth_vs_theta}
\end{figure}

Fig.~\ref{fig:linewidth_vs_theta} shows the measured linewidth as a function of magnetic field angle $\theta_e$. The asymmetric error bars on each data point represent the upper and lower standard deviations of the data set, which is made up of the integrated power at the various current values. The localized mode has its minimum linewidth at $\theta_e\approx30$ deg, while for the propagating mode the linewidth minimum is at $\theta_e\approx60$ deg. When approaching this angle at which the localized mode disappears, the linewidth drops very sharply, by almost two orders of magnitude, as the angle is varied by $\approx2$ deg.

\section{Discussion}
\subsection{Integrated power}
In order to describe the angular dependence of the integrated power, we use a mathematical model based on the evaluation of the nonlinear oscillation power.\cite{PhysRevB.81.184411} Such a quantity, which represents a proper generalization of oscillation power to the case of spatially nonuniform and noncircular magnetization precession, is indeed found to be proportional to the integrated power computed experimentally.\cite{PhysRevB.81.184411,Krivorotov2007}
This model requires knowledge of the closed trajectory of the magnetization vector during precessional motion. By normalizing this vector w.r.t. the saturation magnetization, the dynamics occurs on the unit sphere. The corresponding curve splits the surface of the unit sphere into two parts: a larger part $S_L$, and a smaller part $S_S$. The nonlinear power $P_{N\!L}$ is defined as the sum over all computational cells $N$ of the areas of the smaller surfaces $S_S$:
\begin{equation}
P_{N\!L}=\dfrac{1}{4\pi N}\sum_{j=1}^N\iint_{S_{S_j}}dS
\label{eq:P_NL}
\end{equation}
It should be recalled that, since the simulations performed in Ref.~\onlinecite{consolo:014420} were based on the realistic assumption that the spin-transfer torque only dynamically affects the magnetization configuration of the thinner ``free layer'', the formulation in Eq.~(\ref{eq:P_NL}) will therefore not account for the trajectories described by the magnetization vector of the thicker ``fixed layer''.

\begin{figure}[t!]
\centering
\includegraphics[trim=5cm 8.5cm 5cm 9cm clip=true, scale=0.62]{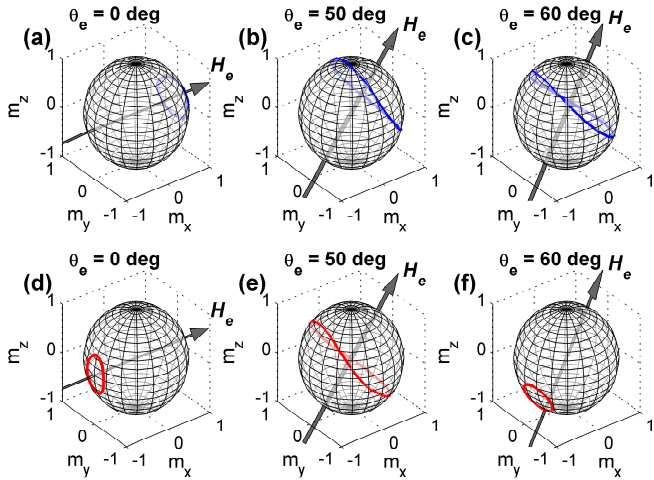}
\caption{(Color online) Precession orbit of the magnetization for the localized mode (top row) and the propagating mode (bottom row), at different applied field directions, indicated by the solid arrow. The magnitude of the field is the same in all cases.}
\label{fig:trajectory}
\end{figure}

The results presented in Fig. \ref{fig:intpow_vs_theta} show that the qualitatively different experimental behaviors [Fig. \ref{fig:intpow_vs_theta}(a)] of the power of the two modes as a function of the applied field angle $\theta_e$ (a monotonic increase for the propagating mode, a relative maximum for the localized mode) can be well reproduced by micromagnetic simulations [Fig. \ref{fig:intpow_vs_theta}(b)]. Quantitative agreements are difficult to obtain for several reasons. Firstly, the discrepancies between the nominal and realistic parameters values are hard to estimate. Secondly, some relevant contributions arising from finite temperature effects, Oersted fields, and nonlinear damping were not considered in the simulations. Recent numerical investigations\cite{PhysRevLett.105.217204} indeed show that a satisfying quantitative agreement between micromagnetic simulations and laboratory experiments can be only achieved in the presence of a large Oersted field. In this framework, it was demonstrated that the inhomogeneities created by this field contribution drastically change the properties exhibited by both the propagating modes (the frequency tunability, for example, can even change sign\cite{PhysRevLett.105.217204}) and localized modes (the threshold current increases\cite{5297537}). One may therefore ask whether simulations that do not include the Oersted field are reliable at all, even if only for qualitative comparison.

For this reason, we performed simulations including the Oersted field at two selected angles $\theta_e = 25$ deg and $\theta_e=70$ deg. At $\theta_e = 25$ deg, we again found the same qualitative behavior as in our recent work,\cite{PhysRevLett.105.217204} namely that the two spin-wave modes are excited alternately. When we compared the precession orbits with and without the inclusion of the Oersted field, we observed that the effect of the Oersted field is one of a slight perturbation. At $\theta_e = 70$ deg, only one mode can be excited and, again, the precession orbit is only slightly changed, which in turn leads to a slight change in the computed power. This is shown in Fig.~\ref{fig:intpow_vs_theta} by open square and filled star symbols for the spin-wave bullet and the propagating mode, respectively. For the spin-wave bullet there is no noticeable change in the computed power, while for the propagating mode the computed power is slightly lower than the case without the Oersted field. The conclusion is therefore that the Oersted field, while having a fundamental role in triggering the switching between the two spin-wave modes, does not perturb in a critical way the trajectory of the magnetization when a particular spin-wave mode is excited. The reason for the discrepancy between the experimental and the computed power could therefore lie in (besides the aforementioned realistic parameter estimation and finite temperature effects) the different dwell times of the two spin-wave modes at different angles. However, it is beyond the scope of this article to investigate such effects.

From the results of the micromagnetic simulations, we are also able to infer the details of the magnetization dynamics. For the propagating mode, the magnetization vector describes a surface on the unit sphere whose area increases monotonically with the increase of the out-of-plane bias field angle $\theta_e$, as shown in Fig.~\ref{fig:intpow_vs_theta}(a)--(c). This leads to the increase in power with $\theta_e$ observed in Fig. \ref{fig:intpow_vs_theta}. Such a result can be understood considering that the demagnetizing field in the thin film is larger at larger $\theta_e$, acting so as to open up the precession cone and to enlarge the corresponding precession surface. This is also confirmed by the fact that, at a given bias current, the precession frequency decreases when $\theta_e$ increases.\cite{PhysRevLett.105.217204,slavinIEEE2005}

Analogous conclusions could be drawn by using a previously-proposed simple model for the integrated power, based on the assumption that the precession orbit is circular.\cite{Rippard2004a} Under this condition, the power $P$ can be related to the maximum GMR signal $\Delta R_{max}$, to the angle between the precessional axis direction in the ``free'' layer and the magnetization in the ``fixed'' layer $\gamma_P$, and to the precession angle $\beta$, through the following relationship:
\begin{equation}
P\sim\Delta R_{max} \sin\gamma_P\sin\beta.
\label{eq:power}
\end{equation}

By performing a similar numerical investigation as a function of $\theta_e$, one can also conclude in this case that the main effect of the increase in the out-of-plane angle is to increase the precessional cone by means of an increase in the angle $\beta$, which in turn determines the increase in the power $P$.

However, it should be pointed out that the linear approach employed in Eq.~(\ref{eq:power}) is not suitable for capturing those details of the magnetization dynamics which are associated with the excitation of nonlinear localized bullet modes. In fact, the closed trajectories described on the unit sphere are, in this case, not circular in the whole range of $\theta_e$. Elliptical clamshell-like curves are described instead. In addition, for large supercriticality, the applied current does not yield a further increase of the precession cone, but rather induces distortions in the shape of the surface. Since the computed output power necessarily takes into account all these mechanisms, the nonlinear oscillation power tool of Eq.~(\ref{eq:P_NL}) appears to be more appropriate.

In this case, the trend observed experimentally for the integrated power can be related to the expansion and contraction of the precession surface described at different out-of-plane angles. In particular, for field direction $\theta_e\approx0$ deg, the magnetization precession takes place about an effective field which is antiparallel to that direction, and the corresponding surface turns out to be relatively small: see Fig. \ref{fig:intpow_vs_theta}(c). The magnetization trajectory undergoes an enlargement as the out-of-plane angle increases up to $\theta_e\approx50$ deg. In such a configuration, the unit vector describes an elliptical quasi-diametric path along the sphere ($S_L\approx S_S$), which yields a maximum in the nonlinear oscillation power: see Fig. \ref{fig:intpow_vs_theta}(d). Any further increase of $\theta_e$ induces a reduction in the precession surface w.r.t. the previous quasi-diametric path. Indeed, the nonlinear power exhibits a rapid decrease down to the critical angle $\theta_e\approx60$ deg: see Fig. \ref{fig:intpow_vs_theta}(e). At this field orientation, the magnetization trajectory associated with localized modes would become null, making excitation of these modes inconceivable. 

Above the critical angle, all the power injected by the current to the system is provided to the propagating mode. 

\subsection{Linewidth and peak power}
In order to properly understand linewidth broadening in STOs one has to include thermal effects, which were neglected in our micromagnetic simulations (where $T=0$ K). Those simulations will therefore not be used in the following. Instead, we rely on two different models which apply to cases where either one or two spin-wave modes can be excited. In the first case, we will use the nonlinear auto-oscillator theory developed in Ref.~\onlinecite{Kim2008PRL100a}, while for the second case we will introduce a new model in which the concept of partial coherence of the oscillator is used.

\subsubsection{Nonlinear auto-oscillator theory}

According to the model in Ref.~\onlinecite{Kim2008PRL100a}, the linewidth of a nonlinear auto-oscillator can be expressed as
\begin{equation}
\Delta f = \dfrac{\Gamma_0}{2\pi}\left(\frac{k_BT}{\beta P_0} \right)\left[1+\left(\frac{N}{\Gamma_{\rm eff}}\right)^2 \right],
\label{eq:lw}
\end{equation}
where $\Gamma_0 = \Gamma_G(1+QP_0)$, $\beta = \mu_0 M_0 \omega_0 V_{\rm eff}/\gamma$, $\Gamma_{\rm eff} = (\alpha_G \omega_0 Q + \sigma I)$, and $\sigma = (\epsilon\gamma\hbar/2eM_0V_{\rm eff})\cos\gamma_p$. $\Gamma_G$ is the Gilbert damping rate (ferromagnetic resonance linewidth), $P_0 = (\zeta-1)/(\zeta+Q)$, where $\zeta=I/I_{th}$, and $Q>0$ is a phenomenological coefficient which quantifies the nonlinearity of the damping.\cite{PhysRevB.75.014440,SlavinTutorial} The explicit expressions for $N$ can be found in Ref.~\onlinecite{SlavinTutorial}. $\gamma_p$ is the angle between the stationary magnetization of the ``free'' and ``fixed'' layers, calculated by solving the magnetostatic boundary conditions. $e$ is the modulus of the electron charge, and $\hbar$ is the reduced Planck constant.

Compared to a conventional auto-oscillator, the extra term involving the $N/\Gamma_{\rm eff}$ ratio (whose value can vary from -$5$ to 30 with our material parameters) in Eq. (\ref{eq:lw}) is the one that governs linewidth broadening in nonlinear auto-oscillators. The angular dependence of the linewidth is also predominantly in the same $N/\Gamma_{\rm eff}$ ratio, since $\Gamma_0$ and $\beta$ are always positive and typically vary by 50\% over the entire angular range. One therefore expects a minimum of the linewidth at the angle where the $N/\Gamma_{\rm eff}$ ratio is minimum. $N$ is a monotonic function of the applied field angle in the range $0^\circ<\theta_e<90^\circ$, and it has a negative value at $\theta_e=0^\circ$, and is positive at $\theta_e=90^\circ$. Therefore, there exists an angle at which $N=0$ and where $\Delta f$ is minimum. This angle coincides with the critical angle $\theta_c$, since the localized mode can be excited only when $N<0$.

In Fig.~\ref{fig:linewidth_vs_theta}, we plot the theoretically calculated linewidth as a continuous line. The numerical constants were set as $\gamma=1.76\times10^{11}$ Hz/T, $\mu_0M_0=0.8$ T, $\alpha_G=0.01$, $V=\pi R^2_c\cdot t$, with $R_c = 20$ nm and $t=4.5$ nm, which are the nominal values for our sample. We also set $Q = (2\omega_M/\omega_0)-1$ according to the theory, and $\zeta=1.5$, which is rather an intermediate value for the currents considered here. The two free parameters are the effective volume $V_{\rm eff}$, which is assumed to be 10 times the volume of a cylinder with the diameter of the nanocontact and height equal to the thickness of the ``free'' magnetic layer, while $\epsilon=0.2$ is chosen as reasonable approximation to the spin-polarization coefficient. From Fig.~\ref{fig:linewidth_vs_theta}, it is evident that the analytical theory correctly predicts the occurrence of the linewidth minimum at the critical angle $\theta_c$. The theory can also describe the data for $\theta_e>\theta_c$, where only one spin-wave mode is excited and the linewidth increases at larger angles. However, the analytical model underestimates the value of the linewidth by almost two orders of magnitude at angles $\theta_e<\theta_c$, where both modes are excited.

\subsubsection{Partial coherence model}
\begin{figure}
\includegraphics[trim=5cm 0cm 5cm 0cm clip=true, scale=0.8]{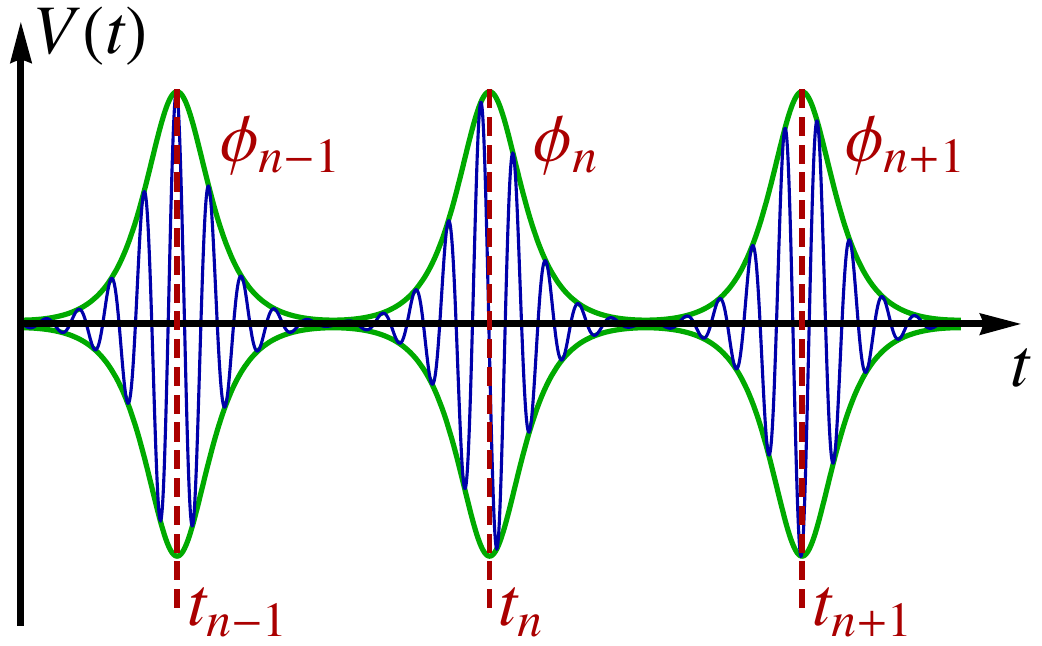}
\caption{Time profile of individual mode oscillations in the bimodal regime of STNO generation (schematically). The time profile consists of a series of ``bursts'' of approximately the same shape at the moments $t_n$. The oscillation phase at each burst is $\phi_n$.}
\label{fig:bursts}
\end{figure}

In order to explain the origin of the generation linewidth (500~MHz) in the bimodal regime of generation, we start from two intuitive considerations, and from from the results of the micromagnetic simulations performed in Ref.~\onlinecite{PhysRevLett.105.217204}.

The first consideration is that we observed hopping between the two modes. Assuming that this process is random (a usual assumption for hopping), the linewidth of each mode would be equal to the inverse of the mode lifetime $\tau\approx0.3$ ns, $\Delta f \sim 1/\tau \approx 3.5$~GHz. Clearly, this interpretation overestimates the linewidth by almost an order of magnitude.

The second consideration is that at 0~K, the hopping between the two modes has a well-defined pattern. That is, the hopping is not random, but is a deterministic periodic process and therefore does not lead to linewidth broadening. The spectrum at 0~K should consist of a series of $\delta$-peaks at multiplies of the hopping frequency $1/T$. At finite temperatures, each spectral peak will be broadened by thermal fluctuations by about the same amount as the linewidth in a single-mode regime. Therefore, the generation linewidth can be estimated using the single-mode theory, $\Delta f \approx 20$~MHz. Clearly, this approach underestimates the linewidth by more than one order of magnitude.

The failure of both the ``completely random'' and ``completely deterministic'' pictures of mode hopping suggests that, in reality the mode hopping is only {\em partially coherent}. That is, the process of persistent hopping happens even at 0~K and is deterministic. However, the power of each mode between ``bursts'' reduces by several orders of magnitude (see Fig.~\ref{fig:bursts}). When the energy of the mode is very low, it becomes much more sensitive to thermal fluctuations (the influence of thermal fluctuations is proportional to $k_BT/E$, where $E$ is the mode energy). As a result, the phase $\phi_n$ of the next burst will not be exactly the same as the phase of the previous burst $\phi_{n-1}$, but will have a significant random component, $\phi_n = \phi_{n-1} + \delta\phi_n$, collected during the low-power stage of the evolution of the mode. In principle, the positions $t_n$ of the bursts' maxima will also be randomized, but we will neglect this effect for simplicity---it has the same qualitative influence on the linewidth as does the randomness of the phases $\phi_n$ (and there are reasons to believe that the randomness in $t_n$ will be smaller than the randomness in $\phi_n$). Thus, here we will assume exact amplitude periodicity: $t_n = n T$.

Under these assumptions, the time profile of the slow (without trivial dependence $e^{-i\omega_0t}$, where $\omega_0$ is the average frequency of oscillations) complex mode amplitude can be written as
\begin{equation}\label{eq-c}
	c(t) = \sum_n a(t-n T)e^{i\phi_n}
\,,\end{equation}
where $a(t)$ describes the amplitude profile of one burst, and $\phi_n$ is the random phase of the $n$-th burst. We will assume that the phase difference $(\phi_n - \phi_{n-1})$ of two consecutive bursts is distributed with probability
\begin{equation}\label{prob}
	P(\phi_n - \phi_{n-1}) = 
			\frac{1}{\sqrt{2\pi} \Delta\phi} \exp\left[-\frac{(\phi_n - \phi_{n-1})^2}{2\Delta\phi^2}\right]
\ .\end{equation}
Here $\Delta\phi$ is the average phase variance between two bursts (phenomenological parameter). The probability distribution Eq.~(\ref{prob}) corresponds to a ``random walk'' of the phase, and is valid for $\Delta\phi < 2\pi$.

Using Eq.~(\ref{eq-c}), one can find the Fourier image of the complex amplitude,
\begin{equation}
	c_\omega = \sum_{n=-\infty}^\infty a_\omega e^{in\omega T + i\phi_n},
\ \end{equation}
where $a_\omega$ is the Fourier image of the amplitude $a(t)$ and the energy spectrum of the oscillations is (note that since we are using a {\em slow} amplitude, the spectrum is measured from the central frequency $\omega_0$)
\begin{equation}\label{S}
	S(\omega) = S_0(\omega) \sum_{k=-\infty}^\infty R_k e^{ik\omega T}
\ .\end{equation}
Here $S_0(\omega) = |a_\omega|^2/T$, $R_k = \left\langle \exp[i(\phi_{n+k} - \phi_n)] \right\rangle$, with the angular brackets denoting averaging over the statistics of phase fluctuations. For the random walk process Eq.~(\ref{prob}), the phase correlator can be easily calculated:
\begin{equation}
	R_k = \exp\left(-\frac{|k|}{2}\Delta\phi^2\right)
\ .\end{equation}
Then, the summation in Eq.~(\ref{S}) can be performed analytically, yielding the final result for the oscillation spectrum:
\begin{equation}\label{spectrum}
	S(\omega) = \coth(\Delta\phi^2/4)\,S_0(\omega)\,\sigma_\phi(\omega)
\,,\end{equation}
where
\begin{equation}\label{spectrum-phi}
	\sigma_\phi(\omega) = \frac{\cosh(\Delta\phi^2/2) - 1}{\cosh(\Delta\phi^2/2) - \cos(\omega T)}
\ .\end{equation}

\begin{figure}
\includegraphics[trim=5cm 0cm 5cm 0cm clip=true, scale=0.8]{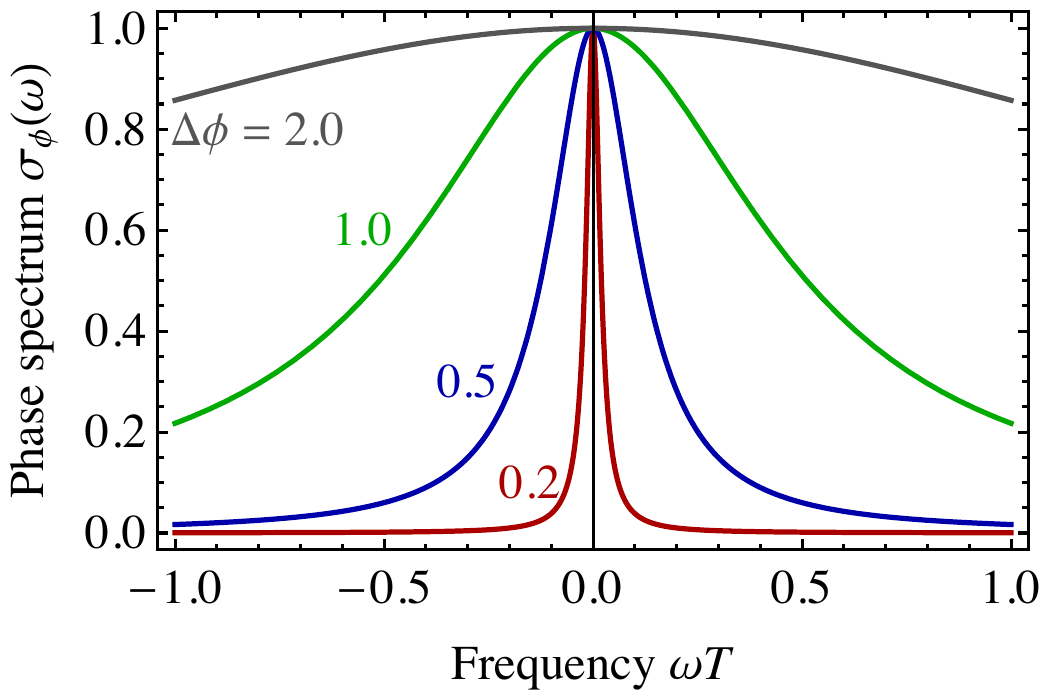}
\caption{Effective spectrum of phase fluctuations $\sigma_\phi(\omega)$ for several values of the phase variance $\Delta\phi$.} 
\label{fig:spectrum_vs_deltaphi}
\end{figure}

The oscillation spectrum $S(\omega)$ is a product of two terms---the spectrum of the individual burst $S_0(\omega)$, and the effective spectrum of phase fluctuations $\sigma_\phi(\omega)$. The spectral width of the burst spectrum $S_0(\omega)$ is of the order of $1/\tau \approx 3.5$~GHz. The spectrum $\sigma_\phi(\omega)$ is shown in Fig.~\ref{fig:spectrum_vs_deltaphi} for several values of the phase variance $\Delta\phi$. One can see that for small $\Delta\phi$ the width of $\sigma_\phi$ is much smaller than $1/T$, whereas for large $\Delta\phi$ it is larger than $1/T$. Two regimes of generation can be distinguished.

In the first (decoherent) regime, $\Delta\phi \gg 1$, different bursts are not correlated, and the spectrum of oscillations is determined by the width of the single-burst spectrum $S_0(\omega)$. This is the standard hopping regime, which is clearly not realized in our case.

In the second (partially coherent) regime, $\Delta\phi \leq 1$, the phases of consecutive bursts are partially correlated, and the generation linewidth is determined by the width of the phase spectrum $\sigma_\phi(\omega)$. Using a series expansion of $\sigma_\phi(\omega)$, one can derive an approximate expression for FWHM linewidth, valid in the limit where $\Delta\phi \ll 1$:
\begin{equation}
	\Delta f = \frac{\Delta\phi^2}{2\pi T}
\ .\end{equation}
Using this expression and the experimental value $\Delta f = 500$~MHz, one can estimate $\Delta\phi = 1.33 = 76^\circ$ (the numerical solution, without series expansion, gives the close value $\Delta\phi = 1.29 = 74^\circ$).

\section{Conclusions}

We measured the power emitted by an NC spin-torque oscillator as a function of out-of-plane applied magnetic field direction. As the angle varies, two distinct spin-wave modes are excited: the Slonczewski propagating mode and the Slavin-Tiberkevich self-localized mode (the spin-wave bullet). The power of the propagating mode increases monotonically as out-of-plane angles are approached, while the power of the localized mode reaches a maximum, and then decreases towards a critical angle, above which the localized mode no longer exists. Both facts are qualitatively reproduced by micromagnetic simulations.

From the micromagnetically simulated nonlinear oscillation power, we were able to infer details of the precession trajectory described by the normalized magnetization vector on the unit sphere. In particular, we found that the trajectories followed by localized modes increase, on average, up to an out-of-plane angle $\theta_e\approx50$ deg (where an almost diametric path is described), and then undergo a rapid decrease down to the critical angle $\theta_c\approx60$ deg (where the magnetization trajectory shrinks towards a single point). Above this critical angle, the excitation of bullet modes is therefore prevented, and all the energy provided by the current to the system only supports the excitation of propagating modes.

Using the nonlinear auto-oscillator model that includes the effect of temperature in the calculation of the generation linewidth, we were able to quantitatively describe the experimental observations when only one spin-wave mode is excited. When two modes are excited, the model underestimates the experimental linewidth by approximately one order of magnitude. Intermediate value of the generation linewidth $\Delta f = 500$~MHz, when both spin-wave modes are excited, could be explained by assuming partial phase coherence between consecutive bursts. The estimated value of the phase variance, $\Delta\phi \simeq 75^\circ$, seems reasonable taking into account the substantial reduction in the power of the mode between the bursts. Since the magnetization dynamics in the bimodal regime is extremely complex, it is unlikely that the value of $\Delta\phi$ can be calculated analytically.

The authors gratefully acknowledge Fred Mancoff at Everspin Technologies for providing the samples, and Giovanni Finocchio for useful discussions. We gratefully acknowledge financial support from the Swedish Research Council (VR), and the Knut and Alice Wallenberg Foundation. Johan \AA kerman is a Royal Swedish Academy of Sciences Research Fellow supported by a grant from the Knut and Alice Wallenberg Foundation.

\end{document}